\documentclass[twocolumn]{aastex631}

\accepted{December 10, 2024}
\submitjournal{ApJ}
\shorttitle{Search for radio MSPs from ellipsoidal ELM WD Binaries}
\shortauthors{Huang et al.}

\usepackage{listings}
\usepackage{bm}
\usepackage{amsmath}
\usepackage{booktabs}
\usepackage{threeparttable}
\usepackage[figuresright]{rotating}

\begin{document}

\title{A Search for Radio Millisecond Pulsar Companions around Extremely Low-mass White Dwarfs with Ellipsoidal Variability}

\correspondingauthor{Pak-Hin Thomas Tam}
\email{tanbxuan@mail.sysu.edu.cn}

\author[0000-0001-5546-8549]{W. J. Huang}
\affiliation{School of Physics and Astronomy, Sun Yat-sen University, Zhuhai 519082, China}

\author[0000-0002-1262-7375]{Pak-Hin Thomas Tam}
\affiliation{School of Physics and Astronomy, Sun Yat-sen University, Zhuhai 519082, China}

\author[0000-0002-1428-4003]{L. L. Ren}
\affiliation{School of Electrical and Electronic Engineering, Anhui Science and Technology University, Bengbu, Anhui 233030, China}

\author[0009-0008-9942-620X]{J. M. Lin}
\affiliation{School of Physics and Astronomy, Sun Yat-sen University, Zhuhai 519082, China}

\begin{abstract}
Extremely low-mass white dwarfs (ELM WDs) are helium-core white dwarfs with masses less than 0.3 $M_{\odot}$. Short-period ELM WD binaries that exhibit ellipsoidal variations may harbor heavier companions, either massive white dwarfs or millisecond pulsars (MSPs). In this study, we selected $\sim$ 12,000 ELM WDs or their candidates, and searched for ellipsoidal-like lightcurves with orbital periods shorter than one day, by using the public data from Zwicky Transient Facility. Finally 23 such systems were found, with 17 being newly discovered. We selected nine high-priority targets likely to evolve from the Roche-lobe overflow channel and estimated their companion masses from the extracted ellipsoidal variation amplitude. Among them, the four targets have companion masses exceeding 1 $M_{\odot}$. We performed a search for radio pulsations from six of these targets by using Five-hundred-meter Aperture Spherical radio Telescope. However, no convincing radio pulsed signals were found, resulting in upper limits for the radio flux at around 8 $\mu$Jy. Given the non-detection of radio pulsations from a total of 11 similar systems, the fraction of ellipsoidal ELM WDs around MSPs is estimated to be below 15$^{+6}_{-3}$\%. We anticipate that multi-wavelength studies of more ellipsoidal-like ELM WDs will further constrain the fraction.
\end{abstract}


\section{Introduction} \label{sec:intro}
    Extremely low-mass white dwarfs (ELM WDs) are commonly considered as helium-core white dwarfs with masses less than $\sim$0.3 $M_{\odot}$. Generally, these systems can exclusively originate within close binaries since the age of the universe hasn't allowed isolated stars to form ELM WDs. The surface hydrogen envelope of the ELM WD's progenitor is stripped away by the companion star, through a stable Roche-lobe overflow (RL) or common envelope ejection (CE) channel \citep{2018ApJ...858...14S,2019ApJ...871..148L}, preventing the conditions necessary for a helium flash and leading to the formation of a low-mass helium-core white dwarf. 

    Most of ELM WDs are discovered in the ELM Survey, which carried out follow-up spectroscopic measurements for ELM WD targets based on photometric selections (e.g. \citealt{2010ApJ...723.1072B,2016ApJ...818..155B,2020ApJ...889...49B,2022ApJ...933...94B}). Currently ELM Survey identified about 150 low-mass WD binaries covering the Southern and Northern Hemispheres and found over a dozen ultracompact WD binaries (orbital period $P_{\rm orb} \leqslant$ 1 hr) which are potential multi-messenger sources that emit gravitational waves at millihertz frequencies \citep{2023ApJ...950..141K}. These are also leading targets detectable by the space-based gravitational wave observatories such as LISA \citep{2018MNRAS.480..302K} and TianQin \citep{2020PhRvD.102f3021H}. Additionally, some studies are dedicated to searching for the more bloated pre-ELM WDs that have not yet begun their cooling track evolution \citep{2021MNRAS.508.4106E,2023AJ....165..119Y}.

    Almost all known ELM WDs are binary systems, and their companions typically encompass millisecond pulsars  (MSPs; \citealt{2014A&A...571A..45I}), type A or F dwarfs (EL CVn–type binaries; \citealt{2014MNRAS.437.1681M}), and, in most instances, white dwarfs \citep{2020ApJ...889...49B}. 
    Generally, MSPs are recycled pulsars that accrete mass and transfer angular momentum from their companions during the phase of being in low-mass X-ray binaries (LMXBs), while most of donors eventually become low mass He-core WDs \citep{1999A&A...350..928T,2005ASPC..328..357V}. At present, the ATNF pulsar catalogue contains over 600 MSPs (spin period $P_{\rm spin}<$ 30 ms), of which approximately 200 are in binary systems with WDs \citep{2005AJ....129.1993M}. By combining the mass function obtained from the MSP spin Doppler effect, it is estimated that the number of MSP/ELM WD systems is roughly 140, assuming a neutron star mass of 1.35 $M_{\odot}$ and an orbital inclination of 60$^{\circ}$. These systems almost all have circular orbits, with orbital periods ranging from 0.05 days to 669 days. It is important to emphasize that the recycled MSP orbiting the WD may be more massive \citep{2020mbhe.confE..23L}, which would result in a greater estimated mass for the companion star.

    In fact, on one hand, the masses of both components in some rare MSP binaries can be accurately estimated through radio timing measurements involving Shapiro delay (e.g., \citealt{2010Natur.467.1081D}) and periastron precession (e.g., \citealt{2008ApJ...679.1433F}). On the other hand, if the WD around MSPs is accessible for detailed spectroscopic measurements, its mass can be derived by comparing the atmospheric parameters with the mass-radius relations of low-mass WDs. The pulsar's mass can then be determined by jointly analyzing the WD's radial velocity variations and the pulsar timing results \citep{2005ASPC..328..357V}.
    
    However, the majority of WDs around MSPs are too faint for optical observations \citep{2005ASPC..328..357V, 2009ApJ...697..283A,2021MNRAS.501.1116A}.
    So far, only about a handful of MSP/ELM WD systems have been confirmed through either radio timing analysis or optical identification,
    such as MSP J0218+4232 \citep{2003A&A...403.1067B}, J0348+0432 \citep{2013Sci...340..448A}, J0437-4715 \citep{2008ApJ...679..675V,2012ApJ...746....6D}, J0614-3329 \citep{2016MNRAS.455.3806B}, J0751+1807 \citep{2016MNRAS.458.3341D}, J1012+5307 \citep{1996ApJ...467L..89V}, J1713+0747 \citep{2005ApJ...620..405S}, J1738+0333 \citep{2012MNRAS.423.3316A}, J1857+0943 \citep{2000ApJ...530L..37V,2018ApJS..235...37A}, J1909-3744 \citep{2005ApJ...629L.113J} and J1911-5958A \citep{2006A&A...456..295B}. 
    Specifically, the first ELM WD was identified spectroscopically as the companion to the PSR J1012+5307 \citep{1996ApJ...467L..89V,2020MNRAS.494.4031M,2024ApJ...962...54W}.  PSR J1738+0333 was the first MSP discovered alongside a pulsating ELM WD companion, exhibiting multimode pulsation periods ranging from 1790 s to 9860 s \citep{2015MNRAS.446L..26K,2018MNRAS.479.1267K}. In addition, since approximately 35\% of MSPs are located in globular clusters,  high-resolution optical telescopes are also highly effective for identifying MSP/ELM WD systems, which can provide valuable insights into the dynamical evolution of MSP binaries (e.g., \citealt{2015ApJ...812...63C,2019ApJ...875...25C}).

    Enlarging the population of compact MSP/ELM WD systems is important to investigate the recycling scenario of MSPs as well as stellar binary evolution (e.g., \citealt{2014A&A...571A..45I,2014A&A...571L...3I,2016A&A...595A..35I}). Particularly, once the ELM WD companions to MSPs are bright enough, their optically-dependent photometric and spectroscopic measurements (such as atmospheric parameters, radial velocity variations, parallax and cooling age) can indirectly improve or constrain estimates for some fundamental parameters of MSPs, including mass, distance, age, and more (e.g., \citealt{2005ASPC..328..357V,2013Sci...340..448A}).  
    Many studies \citep{2007MNRAS.374.1437V,2009ApJ...697..283A,2020ApJ...889...49B,2021MNRAS.505.4981A} have tried to search for radio MSPs around the position of low-mass WDs with high mass functions. However no pulsar signals were detected and the fraction of low-mass WDs orbited by neutron stars (NSs) was estimated to be below 10\% \citep{2021MNRAS.505.4981A}.

    ELM WDs exhibit larger sizes and higher brightness compared to other WDs, due to the inverse correlation between mass and radius \citep{2022ApJ...936....5W}. They are more prone to being tidally distorted by their heavier companions (typically MSPs or CO-core WDs) in close binary systems. This distortion often leads to observable ellipsoidal variability since their projected sizes change as they orbit (e.g. \citealt{2014ApJ...792...39H,2023MNRAS.525.3963K}). Because the amplitude of ellipsoidal variation is roughly proportional to $(M_2/M_1)(R_{1}/a)^{3}$, where $M_1$ and $M_2$ are the masses of the optical primary (ELM WD) and the unseen secondary respectively, $a$ denotes the orbital semi-major axis and $R_{1}$ is the primary's radius \citep{2014ApJ...792...39H}, MSP/ELM WD binaries seem to be more likely to exhibit ellipsoidal variability compared to double WDs in similar orbits. Therefore, searching for tidally distorted ELM WDs can provide one of the clues for finding potential MSPs. 
    Meanwhile, modeling the ellipsoidal variability of ELM WDs can also constrain the system's inclination and companion mass \citep{2014ApJ...792...39H,2018arXiv180905623B}. 

    In this work, we aim to identify ellipsoidal-like lightcurves of ELM WDs with orbital periods shorter than one day, and further analyze whether these systems harbor MSPs through targeted radio observations. The structure of the paper is outlined as follows: In Section \ref{sec:sample_selection}, we introduce the selection of ELM WDs samples. In Section \ref{sec:lightcurve_selection} we describe the identification of ellipsoidal lightcurves. The results will be presented in Section \ref{sec:results}. In Section \ref{sec:discussion}, we discuss the fraction of ellipsoidal ELM WDs containing MSPs, and Section \ref{sec:conclusion} is the final conclusion.

\section{Sample Selection} \label{sec:sample_selection}
    The ongoing Gaia astrometric survey now presents extraordinary opportunities for discovering more faint ELM WDs. Here we adopt three representative catalogues obtained from previous publications based on the analysis of Gaia DR2 and EDR3 data, as our ELM-WD samples or candidates for further study:
    
    (1) Over the course of more than 10 years, the ELM Survey has conducted spectroscopic measurements, obtaining a large dataset of known low-mass WDs as well as their orbital and stellar atmospheric parameters. We select 119\footnote{The orginal sample of low-mass WDs is sourced from the Table 2 in \cite{2020ApJ...889...49B} and Table 1 in \cite{2022ApJ...933...94B}. The number of ELM WDs selected below 0.3 $M_{\odot}$ in these two tables is 105 and 14, respectively. The complete Table 2 in \cite{2020ApJ...889...49B} can be found at \url{https://cdsarc.cds.unistra.fr/viz-bin/cat/J/ApJ/889/49}.} ELM WDs (hereafter referred to as sample S1) with a mass cutoff of 0.3 $M_{\odot}$ from ELM Survey \citep{2020ApJ...889...49B,2022ApJ...933...94B}.
    
    (2) By applying a series of color cuts in the Hertzsprung–Russell diagram with considering the distribution of known ELM WDs and evolutionary model, \cite{2019MNRAS.488.2892P} identified a sample of 5,762 ELM WD candidates\footnote{\url{https://cdsarc.cds.unistra.fr/viz-bin/cat/J/MNRAS/488/2892\#/browse}} (hereafter referred to as sample S2) selected from Gaia DR2. 

    (3) \cite{2021MNRAS.508.3877G} provided a comprehensive catalogue including 1,280,266 WD candidates\footnote{\url{https://warwick.ac.uk/fac/sci/physics/research/astro/research/catalogues/}} from Gaia EDR3 by using some selection criteria. They also estimated the probability to be a white dwarf ($P_{\rm WD}$) for them and derived various stellar parameters, such as effective temperature, surface gravity, and mass.
    To select the high-confidence ELM WDs, we establish the following criteria: (i) all three types of masses ({\ttfamily mass$_{\rm H}$}, {\ttfamily mass$_{\rm He}$}, {\ttfamily mass$_{\rm mixed}$}) derived from different atmospheric models (pure-H models, pure-He and mixed models) are below 0.3 $M_{\odot}$, and (ii) the probability $P_{\rm WD}$ is larger than 0.75. This selection process yields 6,823 ELM WD candidates (hereafter referred to as sample S3). 
    It's worth noting that mass values less than 0.2 $M_{\odot}$ may not be accurate because those parameters are inferred from an invalid range of mass–radius relations \citep{2021MNRAS.508.3877G}.

    After removing overlapping sources from the three samples, we finally obtain a sample comprising $\sim$ 12,000 ELM WDs or their candidates.

\section{Lightcurve Selection} \label{sec:lightcurve_selection}
    \subsection{Periodic Signal Search}
        Zwicky Transient Facility (ZTF; \citealt{2019PASP..131a8002B,2019PASP..131a8003M,2019PASP..131g8001G}) is a time-domain astronomical survey covering the sky north of  $-28^{\circ}$ declination.       
        The employed Palomar 48-inch Schmidt telescope (FoV $\sim$ 47 deg$^{2}$) scans the entire northern sky every two nights in multi-band filters (g,r,i) with limiting magnitude of detection down to $\sim$21. At present, ZTF has been widely applied in the search for periodic lightcurves of WDs (e.g., \citealt{2021MNRAS.508.4106E,10.1093/mnras/stab3293,2023ApJS..264...39R}).
            
        We query lightcurve data from ZTF DR15 for each coordinates within 1 arcsecond. Because ZTF-i data are recorded less frequently, we exclusively analyze sources that have both ZTF-g and ZTF-r data. As a result, nearly 7,000 targets are available for further analyse. The period search range spans from 5 minutes to 1 day, within which there is a higher likelihood of discovering ellipsoidal variations. We employ the Lomb-Scargle (LS) algorithm from {\ttfamily Astropy} \footnote{\url{https://docs.astropy.org/en/stable/timeseries/lombscargle.html}} 
        to search for sinusoidal-like ellipsoidal lightcurves. If the ZTF-g and ZTF-r data display similar maximum LS power in their power spectra, indicating the same most prominent period (``LS period'' thereafter), these sources will be deemed to have reliable periodic signals. 

    \subsection{Lightcurve Identification}
        \begin{figure*}[htpb]
            \centering
            \includegraphics[scale=0.28]{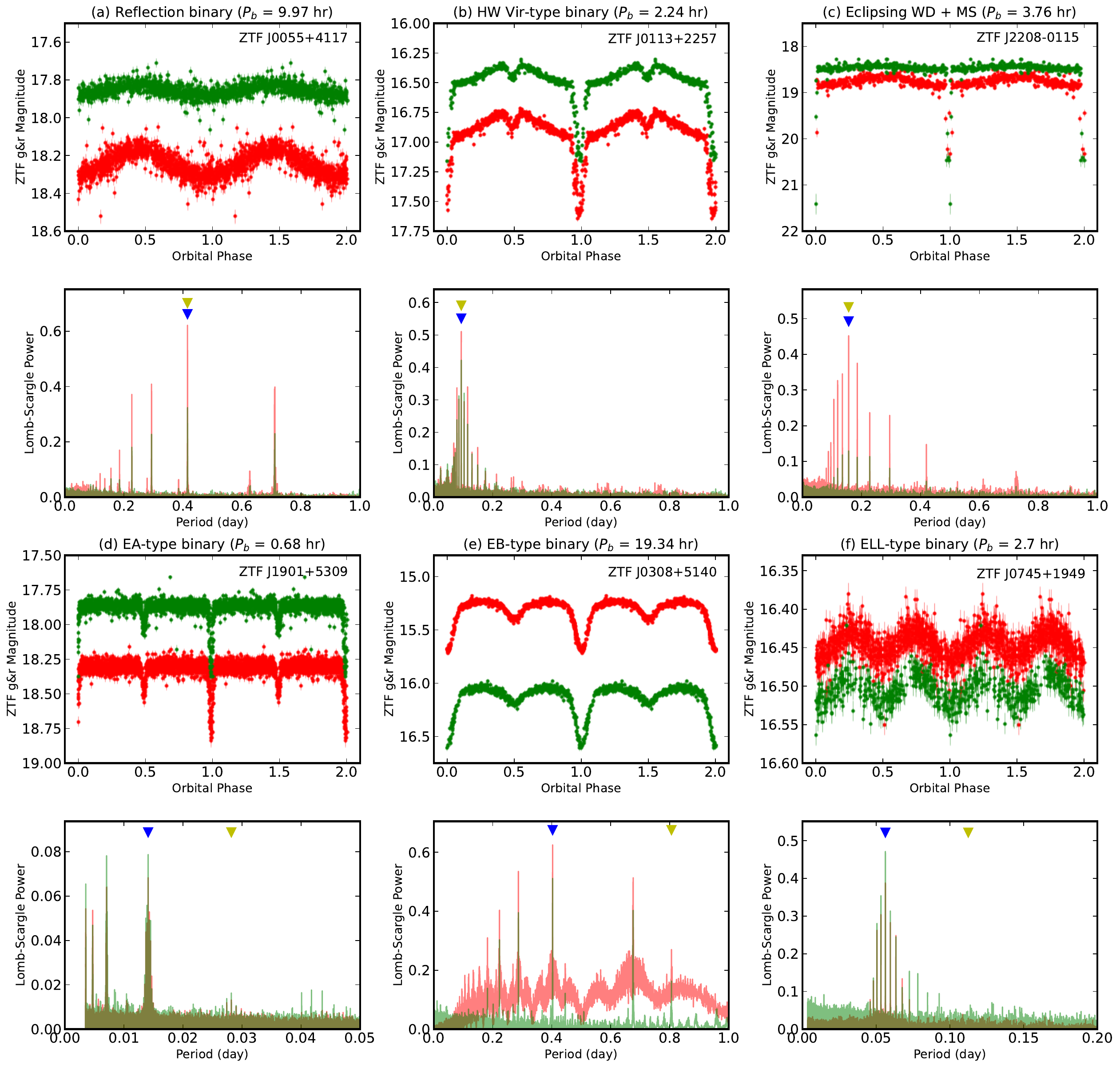}
            \caption{Six primary types of lightcurves (a$-$f) found in our ELM WD samples. ZTF-g and ZTF-r data are represented in green and red, respectively. The panel below each type of lightcurve presents the Lomb-Scargle power spectrums obtained from ZTF-g and ZTF-r data. The blue inverted triangle marks the LS period with the maximum power, while the yellow inverted triangle marks the period at which the lightcurve is actually folded.}
            \label{fig:LCsamples}
        \end{figure*}
        
        During the search process, we encountered many lightcurves with diverse characteristics, and the six primary types are presented in Figure \ref{fig:LCsamples}. They are reflection binaries, ellipsoidal binaries (hereafter ELL-type binaries), and eclipsing binaries, including four subclasses: HW Vir-type binaries, Eclipsing WD+MS systems, EA-type binaries and EB-type binaries. Eclipsing binary systems commonly exhibit extremely sharp dips in the lightcurves, and their shapes depend on the sizes and separation of two stars, their temperatures, the inclination of the orbital plane and other factors. Their orbital periods are typically one or two times the LS period. See \cite{2021MNRAS.508.4106E} and \cite{2023ApJS..264...39R} for more details about these eclipsing systems. In addition, it is unlikely for eclipsing binaries to harbor a NS.
        
        The non-eclipsing reflection binary generally consists of an M dwarf and a hot WD \citep{2021MNRAS.508.4106E}. The M dwarf heated on one side displays a quasi-sinusoidal modulation in its reflected light. Meanwhile, their variability amplitudes in the g-band are notably lower compared to those in the r-band due to a substantial portion of the optical light being contributed by the hot WD. Their orbital periods are directly determined by the LS period.
        
\begin{figure*}[htpb]
            \centering
            \includegraphics[scale=0.35]{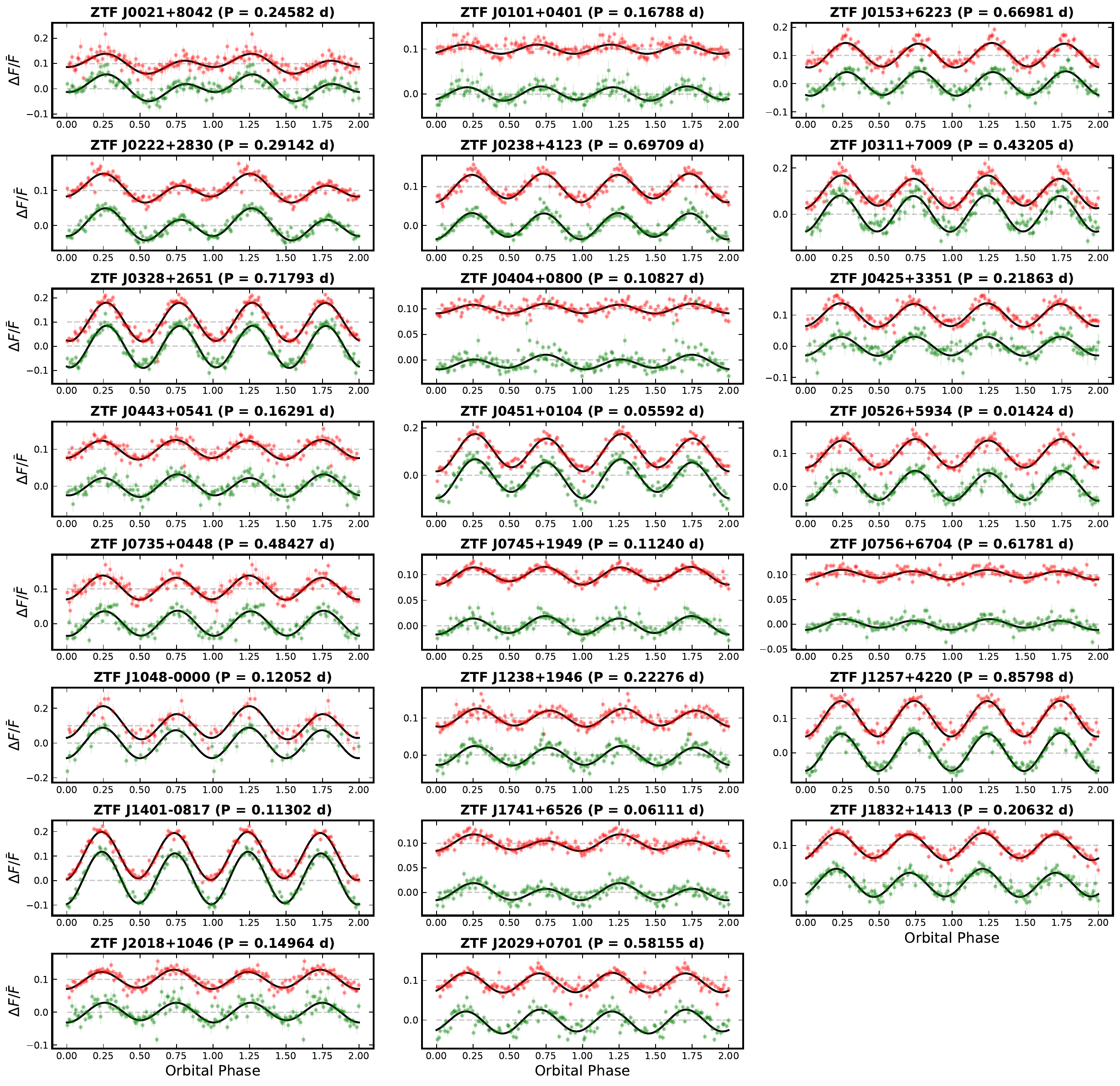}
            \caption{Phase-folded, binned lightcurves in the ZTF-g band (green) and ZTF-r band (red) for 23 ellipsoidal-like variables identified in this work. The r-band lightcurves are vertically shifted upwards by 0.1 for clarity. The black solid lines denote the best-fit curves obtained from the five-parameter model.}
            \label{fig:ELLLC2band}
        \end{figure*}

        \cite{2021MNRAS.508.4106E} suggests the ellipsoidal lightcurves should exhibit double-peak features as well as unequal brightness minima caused by gravity darkening. However, the gravity-darkening effect may not be readily apparent, or can be counteracted by other factors, such as heating from their companions. Similar examples can be seen in other ellipsoidal-like variables \citep{2014ApJ...792...39H,2023MNRAS.525.3963K,2023ApJ...950..141K}, or J0745+1949 showed in Figure \ref{fig:ELLLC2band}.         
        We are not solely aiming to search for pure ellipsoidal variables in order to preserve a greater number of potential samples. Thus alternating brightness minima is excluded in the selection criteria of this work. Of course, this may also involve some interference sources, such as single-mode pulsators and rotators whose lightcurves similar to ellipsoidal variables with equal minima.

        On the other hand, the highest peak in LS power spectrum often identifies half the orbital period of the ELL-type lightcurve, while the power corresponding to the true period is mostly not prominent unless asymmetric double peaks in the lightcurve caused by Doppler beaming effect is present (e.g. \citealt{2014ApJ...792...39H,2022ApJ...933...94B}). Based on this phenomenon, we directly fold the time series at twice the LS period to reconstruct their double-peak structures in the lightcurves.
         
        Due to the absence of alternating minima in the lightcurve and the lack of a distinct peak corresponding to orbital period in LS power spectrum, we mainly distinguish between reflection systems and ellipsoidal variables based on the difference in the amplitude variation of their g-band and r-band lightcurves, since the latter have similar variability amplitudes in both bands. Thus we proceed as follows.

        Firstly, phase-fold lightcurves of double bands are binned into 100 orbital phase points. We assume the epoch ($T_{\rm 0,sup\_conj}$) of the minimum g-band flux as the common zero phase point, corresponding to superior conjunction of the ELM WD. Specifically, for double-peaked lightcurves with different peak heights, possibly due to the Doppler beaming effect, we designate the highest peak as phase 0.25. 
        We then employ a nonlinear least-squares fit that includes an offset and the amplitude of the ellipsoidal variations ($\cos 2\phi$), Doppler beaming ($\sin \phi$), reflection ($\cos \phi$), the first harmonic of the orbital period ($\sin 2\phi$) \citep{2014ApJ...792...39H}.
        Next, we calculate the difference in peak-to-peak amplitudes of double-band lightcurves. Here we define the peak-to-peak amplitude as the difference between the maximum and minimum value of the fitted curve obtained from the five-parameter model. When the difference in peak-to-peak amplitudes of two bands is comparable to the average error of lightcurve data, we consider it to be an ellipsoidal-like variable rather than a reflection system. 
        
        Finally, we obtained 23 ELM WDs or candidates exhibiting ellipsoidal varibility (see Figure \ref{fig:ELLLC2band}), with 7 sources from sample S1, 17 sources from sample S2, and 3 sources from sample S3. Additionally, 17 sources of them are newly discovered to exhibit ellipsoidal variability. Their main parameters are listed in Table \ref{tab:ELLtarget}. 

        \begin{table*}[htpb]
            \tiny
            \centering
            \caption{Astrometric and Atmospheric Parameters of 23 ELL-type Targets}
            \begin{tabular}{lccccccccccc}
                \toprule
                Name & RA & DEC & bp-rp & G & parallax & Period & $T_{\rm eff}$ & $\log g_1$ & $M_{1}$ &  Ref. & Samp.\\
                & (hh:mm:ss.ss)& (dd:mm:ss.s)& (mag) & (mag) & (mas) & (d) & (K) & (cm s$^{-2}$) & ($M_{\odot}$) & & \\
                \hline
                ZTF J0021+8042$^{\ast}$ & 00:21:13.50 & +80:42:35.3 & -0.376 & 17.974 & 1.168$\pm$0.097 & 0.24582 & - & - & - & -  & S2 \\ 
                ZTF J0101+0401$^{\ast}$ & 01:01:28.69 & +04:01:59.0 & 0.156 & 17.293 & 0.796$\pm$0.098 & 0.16788 & $9284^{+120}_{-120}$ & $5.229^{+0.089}_{-0.089}$ & $0.188\pm0.013$ &  [1] & S1,S2 \\ 
                ZTF J0153+6223$^{\ast}$ & 01:53:06.26 & +62:23:19.2 & 0.009 & 17.997 & 2.332$\pm$0.100 & 0.66981 & $23343^{+1835}_{-1835}$ & $7.265^{+0.142}_{-0.142}$ & $0.387\pm0.039$  & [2] & S2\\ 
                ZTF J0222+2830$^{\ast}$ & 02:22:28.37 & +28:30:7.6 & 0.069 & 17.056 & 4.386$\pm$0.114 & 0.29142 & $10485^{+290}_{-290}$ & $6.785^{+0.087}_{-0.087}$ & $0.211\pm0.019$ & [2] & S3\\ 
                ZTF J0238+4123$^{\ast}$ & 02:38:24.05 & +41:23:20.4 & -0.460 & 17.329 & 1.008$\pm$0.101 & 0.69709 & $80210^{+12440}_{-12440}$ & $6.246^{+0.333}_{-0.333}$ & -  & [3] & S2 \\ 
                ZTF J0311+7009$^{\ast}$ & 03:11:10.40 & +70:09:41.9 & 0.013 & 18.311 & 1.557$\pm$0.144 & 0.43205 & - & - & - & -  & S2\\ 
                ZTF J0328+2651$^{\ast}$ & 03:28:03.02 & +26:51:51.6 & -0.229 & 18.179 & 1.524$\pm$0.186 & 0.71793 & $9077^{+311}_{-45}$ & $4.620^{+0.007}_{-0.005}$ & $0.169\pm0.003^{\dagger}$ & [4] & S2\\ 
                ZTF J0404+0800$^{\ast}$ & 04:04:18.39 & +08:00:02.9 & 0.348 & 15.945 & 1.742$\pm$0.044 & 0.10827 & $10860^{+120}_{-120}$ & $4.985^{+0.121}_{-0.121}$ & $0.172\pm0.014$ & [3] & S2\\ 
                ZTF J0425+3351$^{\ast}$ & 04:25:11.52 & +33:51:29.6 & 0.081 & 18.310 & 2.683$\pm$0.188 & 0.21863 & $19653^{+3265}_{-3265}$ & $7.384^{+0.267}_{-0.267}$ & $0.394\pm0.076$ & [2] & S2\\ 
                ZTF J0443+0541$^{\ast}$ & 04:43:02.66 & +05:41:17.0 & 0.297 & 17.289 & 0.760$\pm$0.085 & 0.16291 & $9390^{+140}_{-140}$ & $4.998^{+0.097}_{-0.097}$ & $0.159\pm0.013$ & [3] & S2\\ 
                ZTF J0451+0104 & 04:51:16.84 & +01:04:26.2 & 0.516 & 15.344 & 3.406$\pm$0.033 & 0.05592 & - & - & - & - & S2\\ 
                ZTF J0526+5934 & 05:26:10.42 & +59:34:45.3 & 0.186 & 17.563 & 1.183$\pm$0.091 & 0.01424 & $27300^{+260}_{-260}$ & $6.37^{+0.03}_{-0.03}$ & $0.380\pm0.067$ &  [5] & S2 \\ 
                ZTF J0735+0448$^{\ast}$ & 07:35:29.47 & +04:48:48.8 & -0.328 & 17.570 & 1.055$\pm$0.103 & 0.48427 & $25555^{+3078}_{-3078}$ & $6.475^{+0.242}_{-0.242}$ & $0.227\pm0.028$ & [2] & S2,S3\\ 
                ZTF J0745+1949 & 07:45:11.56 & +19:49:26.6 & 0.384 & 16.406 & 1.088$\pm$0.059 & 0.11240 & $8313^{+100}_{-100}$ & $6.151^{+0.074}_{-0.074}$ & $0.160\pm0.010$  & [6] & S1,S2\\ 
                ZTF J0756+6704$^{\ast}$ & 07:56:10.73 & +67:04:24.4 & 0.035 & 16.410 & 0.484$\pm$0.046 & 0.61781 & 11640$^{+250}_{-250}$ & 4.90$^{+0.14}_{-0.14}$ & 0.181$\pm$0.011 &  [7] & S1\\ 
                ZTF J1048-0000$^{\ast}$ & 10:48:26.86 & -00:00:56.8 & 0.325 & 18.247 & 0.641$\pm$0.195 & 0.12052 & $8484^{+90}_{-90}$ & $5.831^{+0.051}_{-0.051}$ & $0.169\pm0.016$ &  [6] & S1\\
                ZTF J1238+1946$^{\ast}$ & 12:38:00.08 & +19:46:31.1 & -0.144 & 17.487 & 0.452$\pm$0.099 & 0.22276 & $14950^{+420}_{-420}$ & $4.890^{+0.050}_{-0.050}$ & $0.197\pm0.012$ & [6] & S1,S2\\ 
                ZTF J1257+4220 & 12:57:23.99 & +42:20:53.5 & -0.401 & 17.442 & 1.154$\pm$0.109 & 0.85798 & $43974^{+347}_{-347}$ & $7.612^{+0.037}_{-0.037}$ & $0.524\pm0.003$ & [8] & S3 \\ 
                ZTF J1401-0817 & 14:01:18.80 & -08:17:23.5 & 0.243 & 16.663 & 0.823$\pm$0.079 & 0.11302 & $8813^{+90}_{-90}$ & $5.731^{+0.048}_{-0.048}$ & $0.216\pm0.042$ &  [6] & S1 \\ 
                ZTF J1741+6526 & 17:41:40.49 & +65:26:38.6 & 0.145 & 18.456 & 0.867$\pm$0.143 & 0.06111 & $10540^{+170}_{-170}$ & 6.000$^{+0.060}_{-0.060}$ & $0.170\pm0.010$ & [6] & S1 \\
                ZTF J1832+1413$^{\ast}$ & 18:32:45.51 & +14:13:11.3 & 0.589 & 17.454 & 0.665$\pm$0.092 & 0.20632 & $6862^{+49}_{-17}$ & $4.473^{+0.010}_{-0.010}$ & - & [4] & S2\\ 
                ZTF J2018+1046$^{\ast}$ & 20:18:9.99 & +10:46:47.1 & 0.459 & 17.436 & 0.829$\pm$0.089 & 0.14964 & $8216^{+115}_{-154}$ & $4.427^{+0.042}_{-0.075}$ & $0.167\pm0.005^{\dagger}$ & [4] & S2\\ 
                ZTF J2029+0701$^{\ast}$ & 20:29:32.52 & +07:01:7.7 & -0.411 & 16.624 & 1.915$\pm$0.068 & 0.58155 & $63102^{+16916}_{-16916}$ & $7.437^{+0.172}_{-0.172}$ & $0.519\pm0.047$ & [2] & S2\\ 
                \bottomrule
            \end{tabular}
                \raggedright\footnotesize{\textbf{Note.} The source name with ``$\ast$'' indicates it's a newly discovered ellipsoidal-like variable in this paper. The values of coordinates (RA,DEC), BP-RP colour (bp-rp), G-band mean magnitude (G) and parallax are provided from Gaia DR3. The orbital periods are obtained from the ZTF lightcurves. The effective temperature ($T_{\rm eff}$), surface gravity ($\log g_{1}$) and WD mass ($M_{1}$) are directly collected from existing publications or Gaia DR3, showed in Ref. column. An exception is that the masses of two WDs, marked with ``$\dagger$'', are derived from \cite{2013A&A...557A..19A}, through bilinear interpolation of the $T_{\rm eff}$ and $\log g_{1}$ given by Gaia DR3 which may not be entirely accurate. The column of Samp. indicates the sample from which each source originates. \\
                \textbf{Ref.} ~[1] \cite{2022ApJ...933...94B} ~[2] \cite{2021MNRAS.508.3877G} ~[3] \cite{2022ApJ...936....5W} ~[4] Gaia DR3 ~[5] \cite{2023ApJ...950..141K} ~[6] \cite{2020ApJ...889...49B} ~[7] \cite{2015ApJ...812..167G} ~[8] \cite{2019MNRAS.486.2169K}.
                }
            \label{tab:ELLtarget}
        \end{table*}

\section{Results} \label{sec:results}
    \subsection{Distribution in CMD}    
\begin{figure}[htp]
            \centering
            \includegraphics[scale=0.4]{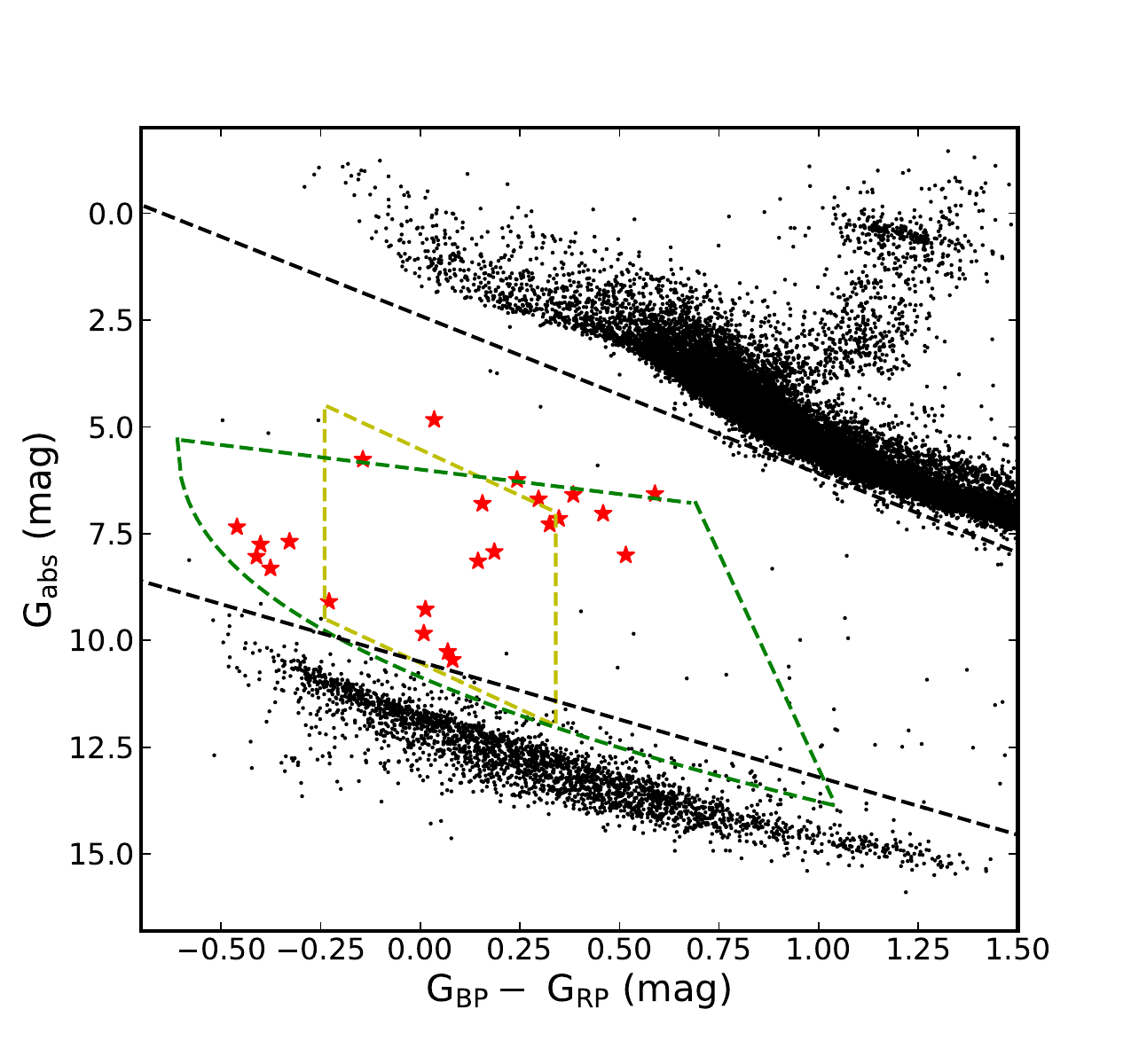}
            \caption{Distribution of 23 ELM WDs or candidates (marked by red stars) on the CMD. Black scatter points show the Gaia stars within 200 pc. The yellow dashed lines delineate the region where ELM WDs identified in the ELM Survey are predominantly located \citep{2023ApJ...950..141K}. The green dashed lines represent the selection region employed to search for ELM WD candidates based on Gaia DR2 data \citep{2019MNRAS.488.2892P}. Low-mass white dwarf companions in MSP binaries are more likely to be found within two black dashed lines, suggested by \cite{2021MNRAS.501.1116A}.}
            \label{fig:HRdiagram}
        \end{figure}

        Figure \ref{fig:HRdiagram} depicts the distribution of 23 ELM WDs or their candidates in Gaia color-magnitude diagram (CMD). They all fall between the main sequence stars and WDs, within the region recommended by \cite{2021MNRAS.501.1116A} for locating low-mass WDs companions to MSPs. For comparison, the search region of the ELM survey \citep{2023ApJ...950..141K} and the region suggested by \cite{2019MNRAS.488.2892P} for finding ELM WDs are also plotted on the CMD. The former represents the distribution of most known ELM WDs to date, while the irregular region of the latter can effectively distinguish ELM WD candidates from the surrounding higher mass WDs, hot subdwarfs, and scattered main sequence stars. The majority of our sources also fall within these regions. 
        
    \subsection{High-priority ELM WD Candidates} \label{sec:cleanSample}
        The effective temperature and surface gravity of the sample listed in Table \ref{tab:ELLtarget} span a wide range, with some candidates likely belonging to sdA stars ( $\log g_{1} \sim$ 6 and $T_{\rm eff}$ $<$ 9,000 K) \citep{2017ApJ...839...23B} or sdB stars (25,000 K $<$ $T_{\rm eff}$ $<$ 40,000 K and 5 $<$ $\log g_1$ $<$ 6) \citep{2009ARA&A..47..211H}. \cite{2020ApJ...889...49B} employed a relatively narrow selection criterion with 8,800 K $\lesssim T_{\rm eff} \lesssim$ 22,000 K and $5.5 \lesssim \log g_1 \lesssim 7.1$ to identify clean samples of ELM WDs. However, under this criterion, only three sources in our sample meet the requirements. To increase the number of ELM WD candidates for further analysis, we select objects with well-measured atmospheric parameters, whose effective temperatures and surface gravities fall within the broader ranges of 8,000 K $\lesssim T_{\rm eff} \lesssim$ 22,000 K and $5 \lesssim \log g_1 \lesssim 7$, as recommended by \cite{2016ApJ...818..155B} for the empirical range of ELM WDs. The total of nine selected targets is listed in Table \ref{tab:ELLtarget2}. 

\begin{figure}
            \centering
            \includegraphics[width=1.1\linewidth]{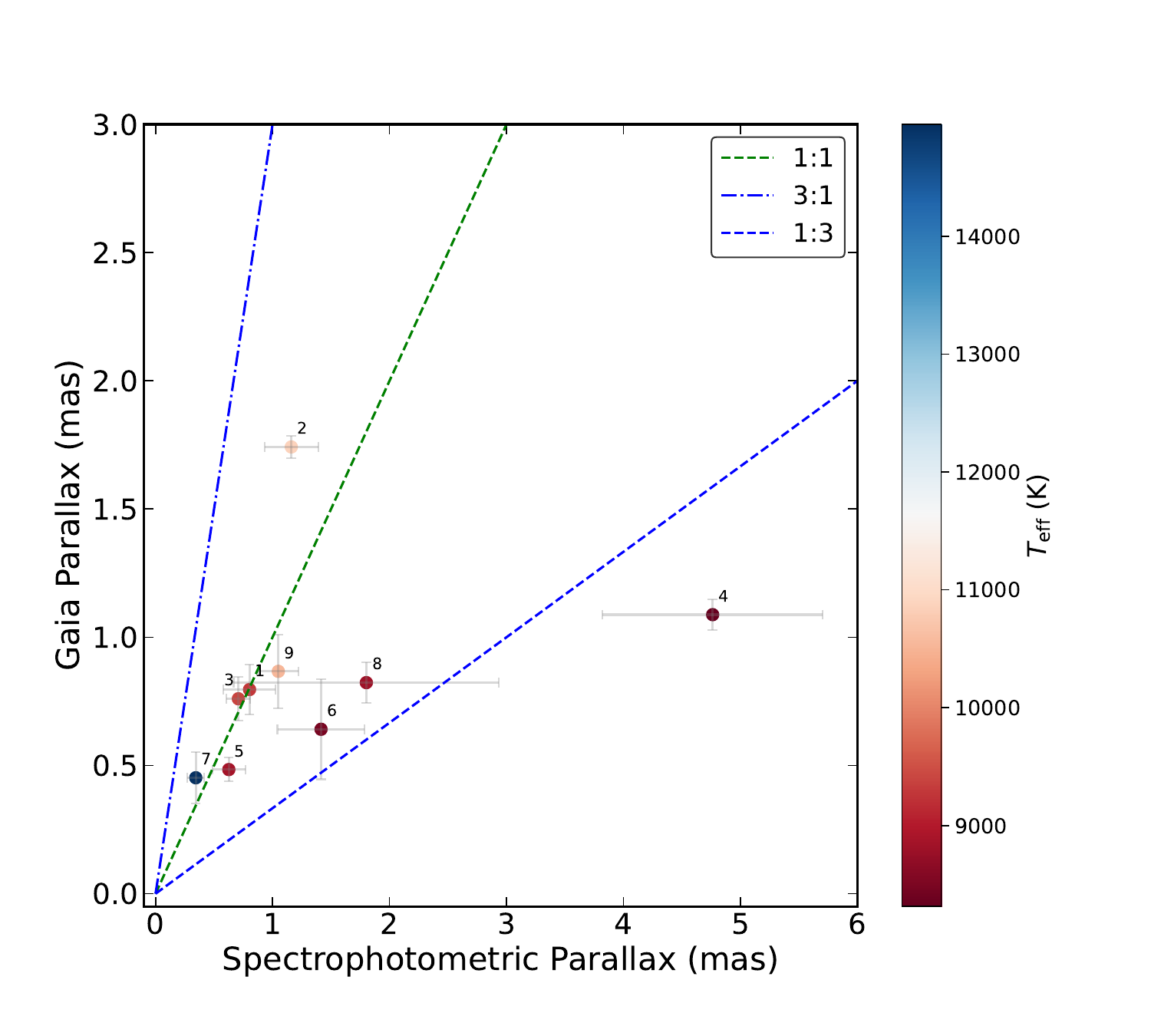}
            \caption{Spectrophotometric parallax estimates versus Gaia parallaxes. The green dashed line denotes the 1:1 ratio line, while the blue dashed and dashed-dotted lines represent the 1:3 and 3:1 ratio lines, respectively. The color of the dots indicates different effective temperatures, and the numerical labels correspond to the source numbers in Table \ref{tab:ELLtarget2}.}
            \label{fig:parallaxVSspecparallax}
        \end{figure}

        We then carefully compared their spectrophotometric parallaxes, inferred from the evolutionary tracks of ELM WDs, with the Gaia parallaxes. If the difference between the two does not exceed a factor of three, the object can be considered a high-priority ELM WD candidate \citep{2020ApJ...889...49B,2022ApJ...936....5W}. The comparison results are shown in Figure \ref{fig:parallaxVSspecparallax}. As seen, except for J0745+1949, all other sources fall between the 1:3 ratio line and the 3:1 ratio line, and can be considered good candidates. 

        \begin{table*}[htpb]
                \scriptsize
                \centering
                \caption{The Measured and Derived Parameters of the 9 High-priority ELM WDs}
                \begin{tabular}{cccccccccc}
                    \toprule
                    No. & Name & RA & DEC & $T_{\rm eff}$ & $\log g_1$ & $M_{1}$ &  Parallax$_{\rm Gaia}$ & Parallax$_{\rm spec}$ & Distance$_{\rm spec}$\\
                    & & (hh:mm:ss.ss)& (dd:mm:ss.s)& (K) & (cm s$^{-2}$) & ($M_{\odot}$) &  (mas) & (mas) & (kpc)\\
                    \hline
                    1& ZTF J0101+0401 & 01:01:28.69 & +04:01:59.0 & $9284^{+120}_{-120}$ & $5.229^{+0.089}_{-0.089}$ & $0.188\pm0.013$  & 0.796$\pm$0.098 & 0.803$\pm$0.222 & 1.245$\pm$0.321 \\
                    2&ZTF J0404+0800 & 04:04:18.39 & +08:00:02.9 & $10860^{+120}_{-120}$ & $4.985^{+0.121}_{-0.121}$ & $0.172\pm0.014$  & 1.742$\pm$0.044 & 1.160$\pm$0.229 & 0.862$\pm$0.177 \\ 
                    3&ZTF J0443+0541 & 04:43:02.66 & +05:41:17.0 & $9390^{+140}_{-140}$ & $4.998^{+0.097}_{-0.097}$ & $0.159\pm0.013$  & 0.760$\pm$0.085 & 0.708$\pm$0.104 & 1.413$\pm$0.211 \\ 
                    4&ZTF J0745+1949 & 07:45:11.56 & +19:49:26.6 & $8313^{+100}_{-100}$ & $6.151^{+0.074}_{-0.074}$ & $0.160\pm0.010$   & 1.088$\pm$0.059 & 4.762$\pm$0.941 & 0.210$\pm$0.040 \\ 
                    5&ZTF J0756+6704 & 07:56:10.73 & +67:04:24.4 & $11640^{+250}_{-250}$ & $4.90^{+0.14}_{-0.14}$ & $0.181\pm0.011$ & 0.484$\pm$0.046 & 0.626$\pm$0.141 & 1.597$\pm$0.343 \\ 
                    6&ZTF J1048-0000 & 10:48:26.86 & -00:00:56.8 & $8484^{+90}_{-90}$ & $5.831^{+0.051}_{-0.051}$ & $0.169\pm0.016$ & 0.641$\pm$0.195 & 1.414$\pm$0.373 & 0.707$\pm$0.175 \\
                    7&ZTF J1238+1946 & 12:38:00.08 & +19:46:31.1 & $14950^{+420}_{-420}$ & $4.890^{+0.050}_{-0.050}$ & $0.197\pm0.012$ & 0.452$\pm$0.099 & 0.344$\pm$0.072 & 2.908$\pm$0.583 \\ 
                    8&ZTF J1401-0817 & 14:01:18.80 & -08:17:23.5 & $8813^{+90}_{-90}$ & $5.731^{+0.048}_{-0.048}$ & $0.216\pm0.042$ & 0.823$\pm$0.079 & 1.802$\pm$1.135 & 0.555$\pm$0.268  \\ 
                    9&ZTF J1741+6526 & 17:41:40.49 & +65:26:38.6 & $10540^{+170}_{-170}$ & $6.000^{+0.060}_{-0.060}$ & $0.170\pm0.010$ & 0.867$\pm$0.143 & 1.049$\pm$0.171 & 0.953$\pm$0.151  \\
                    \bottomrule
                \end{tabular}
                \raggedright\footnotesize{\textbf{Note.} The values for spectrophotometric parallax (Parallax$_{\rm spec}$) or spectrophotometric distance (Distance$_{\rm spec}$) are taken from the references of the corresponding sources listed in Table \ref{tab:ELLtarget}.}
                \label{tab:ELLtarget2}
            \end{table*} 

        The two cooler ($\textless$ 9,000 K) ELM WDs, J1048-0000 and J1401-0817, fall below the 1:1 ratio line and exhibit relatively large deviations, somewhat similar to sdA-type stars. This can be interpreted as either the inflation of the radii or systematic errors in surface gravity \citep{2017ApJ...839...23B,2020ApJ...889...49B}. \cite{2021MNRAS.508.4106E} argued that the $\log g_1$ for J1401-0817 should be 4.93$\pm$0.1, rather than the overestimated reported value ($\sim$ 5.731), in order to account for the observed significant amplitude of ellipsoidal variability. The situation for J1048-0000 is likely similar. If their actual $\log g_1$ values are indeed smaller than the reported ones, their larger radii would bring the spectrophotometric parallax estimates closer to the Gaia parallaxes. The estimated parallax of J0745+1949 shows the largest deviation from the Gaia parallax, suggesting that its mass or radius, as derived from the ELM WD evolutionary model, may be incorrect. In fact, J0745+1949 exhibits abundant metal lines, indicating that it may have recently experienced a CNO flash and might not yet be on the final cooling track \citep{2014ApJ...781..104G,2014ApJ...792...39H}. This characteristic is somewhat similar to that of the companion of PSR J1816+4510, which is believed to be a low mass proto-WD \citep{2012ApJ...753..174K,2013ApJ...765..158K,2014A&A...571L...3I}. Therefore, we retain J0745+1949 as a ``high-priority'' candidate for investigating the potential presence of a pulsar through studies at other wavelengths. Notably, J0756+6704 is a pre-ELM WD exhibiting p-mode pulsations with periods of 521 s and 587 s \citep{2016ApJ...822L..27G}. Its $T_{\rm eff}$ and $\log g_1$ differ from those derived by \cite{2020ApJ...889...49B}, whose estimated parallax ($\sim$1.443) shows a notable deviation from the Gaia parallax.

        We also investigate the possible counterparts of other non-high-priority sources listed in Table \ref{tab:ELLtarget}, although most still require detailed spectroscopic measurements for further confirmation of their nature. J0153+6223, J0222+2830, J0425+3351, and J0735+0448 are high-probability ($P_{\rm WD}>0.75$) low-mass WD candidates in the WD catalogue of \cite{2021MNRAS.508.3877G}. It is worth noting that the $T_{\rm eff}$ and $\log g_1$ of J0222+2830 also meet the criteria for an ELM WD, and its spectrophotometric parallax is approximately 4.652$\pm$0.577, as estimated using the given atmospheric parameters and the method described by \cite{2022ApJ...936....5W}. This value is very close to the Gaia parallax ($\sim$4.386). By checking the Gaia DR3 source ID in SIMBAD\footnote{\url{https://simbad.u-strasbg.fr/simbad/}}, it may also correspond to KUV 02196+2816, which is considered to be a double-degenerate WD system containing a DA star and a DB star \citep{2009ApJ...696.1461L}.
    
        J2029+0701 is also a high-probability WD, although its high temperature ($\sim$ 63,000 K) and $\sim$ 0.5 $M_{\odot}$ mass seem to correspond to a canonical sdB-type star. \cite{2021MNRAS.508.3877G} pointed out that parameters above 40,000 K are unreliable and require further spectroscopic observations for confirmation. J1257+4220 was classified as a hot DA-type WD with UHE lines, and it is possibly a binary system \citep{2019MNRAS.486.2169K,2021A&A...647A.184R}. J0238+4123 has the highest effective temperature ($\sim$ 80,000 K), making it seem to be a hot sdB-type star. 
        The photometric variability of J0451+0104 and J0526+5934 was first discovered by \cite{2023ApJS..264...39R}, and they are potential gravitational wave candidates with orbital periods under 100 minutes. 
        In particular, J0526+5934, with an orbital period of only 40 minutes, is identified as a low-mass WD or post-core-burning hot subdwarf, accompanied by a massive CO-core WD \citep{2023ApJ...959..114K,2024A&A...686A.221R}. The remaining five either have missing atmospheric parameters or parameters provided only by Gaia DR3, necessitating further spectroscopic measurements and studies for identification.

    \subsection{Mass–Period Distribution} \label{sec:mass2period}
        ELM WDs may form through either the CE channel or the RL channel, as suggested by binary evolution theory. However, it is improbable that ELM WDs with MSPs are formed via the CE channel, as NSs are unable to accrete sufficient material to become MSPs \citep{2019ApJ...871..148L}. The mass-period diagram provides a clear reflection of the formation channel of ELM WDs.

        \begin{figure}[htpb]
            \centering
            \includegraphics[scale=0.48]{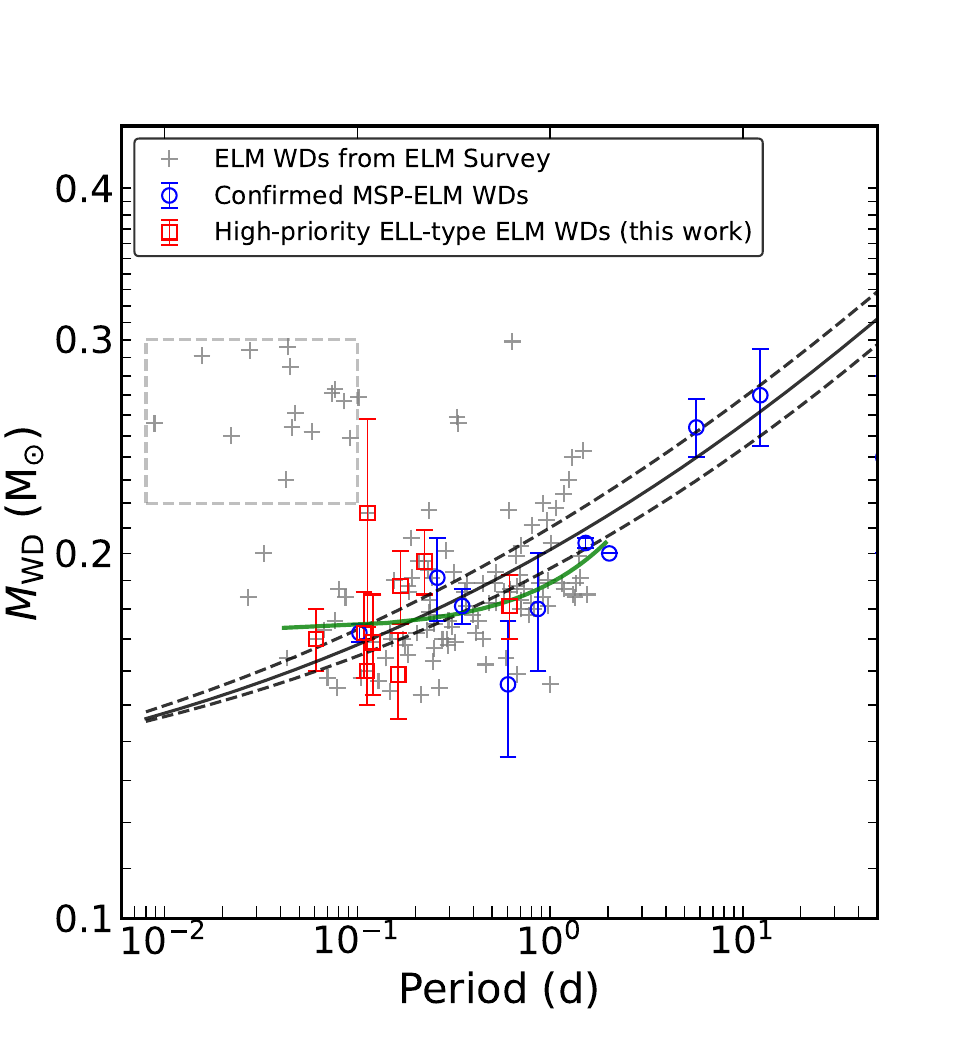}
            \caption{Mass-period distribution of ELM WDs. Nine high-priority ELL-type ELM WDs of our sample are plotted by red squares. ELM WDs from sample S1 are depicted by gray crosses. The gray rectangular box marks the region corresponding to ELM WDs that should follow the CE channel. The green solid line represents the best-fit curve based on a linear function, for the gray crosses outside of the gray rectangle. Some confirmed MSP/ELM WDs are denoted by blue circles, and they conform well to the relationships predicted in \cite{1999A&A...350..928T}, showed by solid and dashed lines with different chemical composition (Pop.I, Pop.I+II and Pop.II).}
            \label{fig:p2m1}
        \end{figure}

        Figure \ref{fig:p2m1} illustrates the distribution of nine high-priority ELL-type ELM WDs on the mass-period diagram, shown by red square. Some confirmed ELM WDs orbiting with MSPs\footnote{These confirmed MSP-ELM WDs are obtained in the Table A1 from \cite{2019MNRAS.488.2892P}. We exclude the PSR J1959+2048, a black widow system containing a 0.035 $M_{\odot}$ companion. The very low mass of the companion may be attributed to ablation from the strong pulsar wind, which seems to be different from the population of normal MSP-WD systems. We also exclude PSR J2317+1439 and PSR J1853+1303 due to their excessively large error bars for the companion mass.} are displayed by blue circles. Their distribution broadly conforms to the relationship predicted by \cite{1999A&A...350..928T}, depicted by the black solid and dash lines representing the different chemical composition of the donor star.
        We also present the ELM WDs from sample S1, plotted as gray cross on the diagram, which are commonly considered to be double WDs. This group of ELM WDs can be divided into two parts \citep{2019ApJ...871..148L,2020ApJ...889...49B}: the ELM WDs within the gray dashed box (0.22 $M_{\odot}$ $\textless$ $M_{\rm WD}$ $\textless$ 0.3 $M_{\odot}$, $P_{\rm orb}$ $\textless$ 0.1 d) may have evolved from the CE channels, which produce more massive He WDs (0.21$-$0.4 $M_{\odot}$) along with very short orbits; the remaining sources show positive correlation between mass and period, likely consistent with the RL channels. The green solid line shows the best-fit curve based on a linear function, with slope and intercept values of 0.023(3) and 0.171(2), respectively. It can be seen that our nine high-priority ELM WDs roughly follow the evolutionary model of the RL channel or \cite{1999A&A...350..928T}.

    \subsection{Estimation of the Companion Mass}\label{subsec: M2}
        We select nine high-priority ELL-type ELM WDs listed in Table \ref{tab:ELLtarget2}, and the mass of their unseen companions can be further estimated from the amplitude of ellipsoidal variations. The ZTF g-band or r-band amplitude of ellipsoidal variation ($\cos2\phi$ term) has been separated in the lightcurve fitting based on five-parameter model in Section \ref{sec:lightcurve_selection}. Here we only analyze the ellipsoidal variation amplitude in the g-band. This amplitude, $A_{\rm EV}$, is principally determined by \citep{2014ApJ...792...39H,1993ApJ...419..344M}:
        \begin{equation}
            A_{\rm EV} = \frac{3\pi^{2}(15+\mu_{1})(1+\tau_{1})M_{2}R_{1}^{3}\sin^{2}i}{5P_{\rm orb}^{2}(3-\mu_{1})GM_{1}(M_{1}+M_{2})},
            \label{eq:ELL}
        \end{equation}
        where $M_{1}$ and $R_{1}$ are the mass and radius of the ELM WD, $M_2$ is the mass of unseen companion, and $i$ is the orbital inclination angle. $\mu_{1}$ and $\tau_{1}$ are linear limb-darkening coefficient and gravity-darkening coefficient of the primary, respectively. Here we interpolate $\mu_{1}$ from \cite{2020A&A...634A..93C}, according to given $T_{\rm eff}$ and $\log g_1$ in Table \ref{tab:ELLtarget}. The reference values from SDSS-g$^{\prime}$ filter are employed, as it shares a similar central wavelength with the ZTF g-band. The temperature-dependent $\tau_{1}$ are calculated by the Equation (10) from \cite{1985ApJ...295..143M}.

        The radius $R_{1}$ can be straightforwardly estimated using the known surface gravity $g_{1}$ and primary mass $M_{1}$:
        \begin{equation}
            R_{1}^{2} = \frac{GM_{1}}{g_{1}}.
            \label{eq:g1}
        \end{equation}
        
        Additionally, the mass function $f_{1}({\rm M2})$ derived from spectroscopy measurements is expressed as follow:
        \begin{equation}
            f_{1}(M_2) = \frac{P_{\rm orb}K_{1}^{3}}{2\pi G} = \frac{M_{2}^{3}\sin^{3}i}{(M_{1}+M_{2})^{2}},
            \label{eq:mass_func}
        \end{equation}
        where $K_{1}$ is the radial velocity amplitude of the optical primary.
        
        Therefore, solving Equation (\ref{eq:ELL}) and (\ref{eq:mass_func}) simultaneously enables us to determine the values of the two unknowns: companion mass $M_{2}$ and inclination angel $i$. 
        
\begin{figure}[ht]
            \centering
            \includegraphics[scale=0.33]{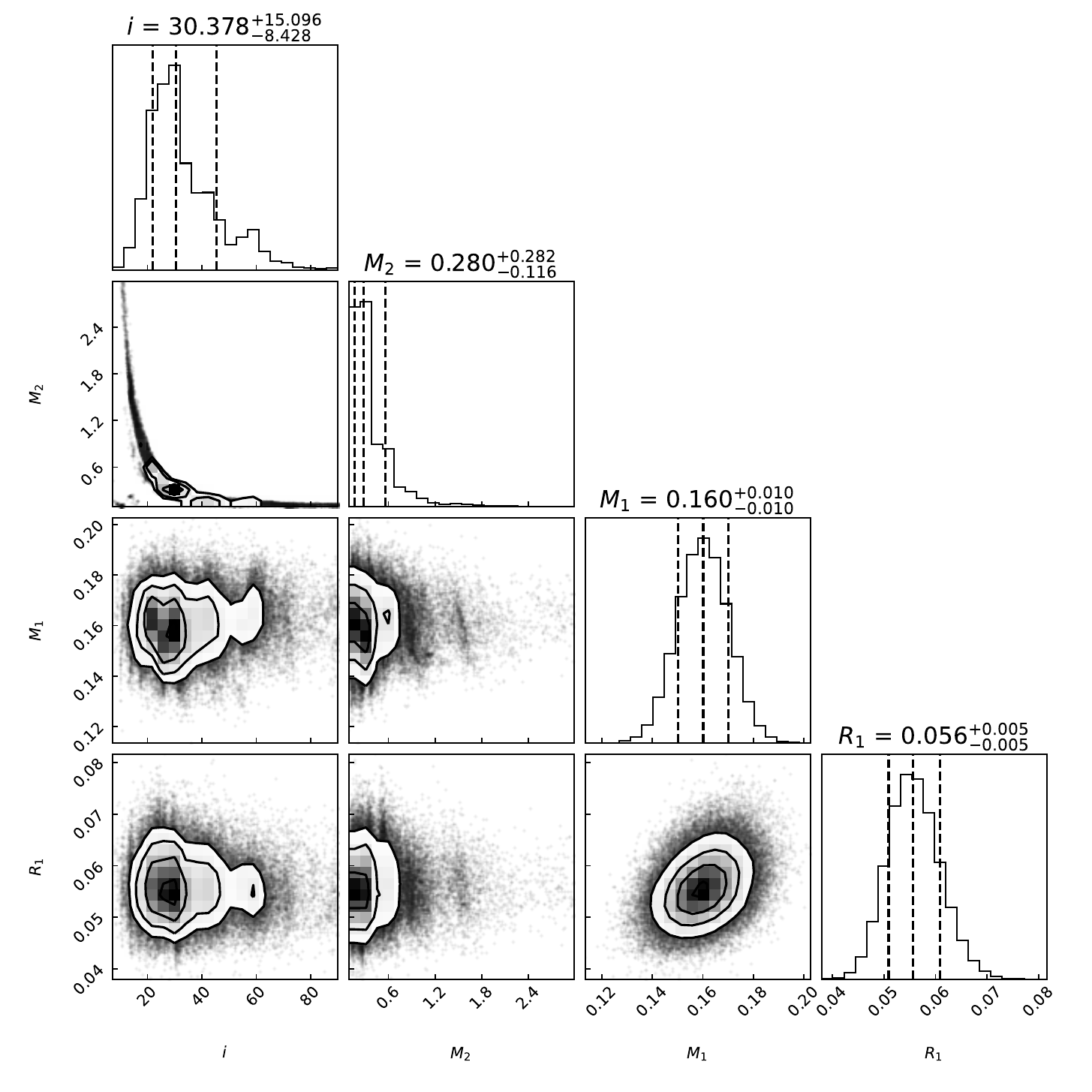}
            \caption{MC results of J0745+1949 for the joint distributions of four parameters ($M_{1}$, $R_{1}$, $M_{2}$ and $i$). The middle vertical dashed lines on the histograms indicate the median value, while the other two dashed lines represent the 1$\sigma$ level.}
            \label{fig:MCresults}
        \end{figure}

        In order to determine the effective range of values for $M_{2}$ and $i$, we perform 100,000 Monte Carlo (MC) simulations to numerically solve equations for 7 targets with derived mass functions. For each simulation, the orbital period is fixed, while $A_{\rm EV}$, $M_{1}$, $g_{1}$ and $f_{1}(M_2)$ are randomly chosen from a Gaussian distribution within their respective measured uncertainties. The minimum mass of the secondary, $M_{\rm 2,min}$, can be inferred form $f_{1}({\rm M2})$ by assuming an inclination angel of 90$^{\circ}$. Meanwhile, the secondary is expected to be a WD or NS with mass less than 3 $M_{\odot}$.
        So we only accept the solutions that have $M_{\rm 2,min}<M_{2} <3~M_{\odot}$ and $M_{1} <$ 1.4 $M_{\odot}$. For instance, Figure \ref{fig:MCresults} illustrates  one- and two-dimensional projections for the valid solutions of J0745+1949. An obvious degeneracy exists between the companion mass $M_{2}$ and the inclination angle $i$. 

        \begin{figure}[ht]
            \centering
            \includegraphics[scale=0.3]{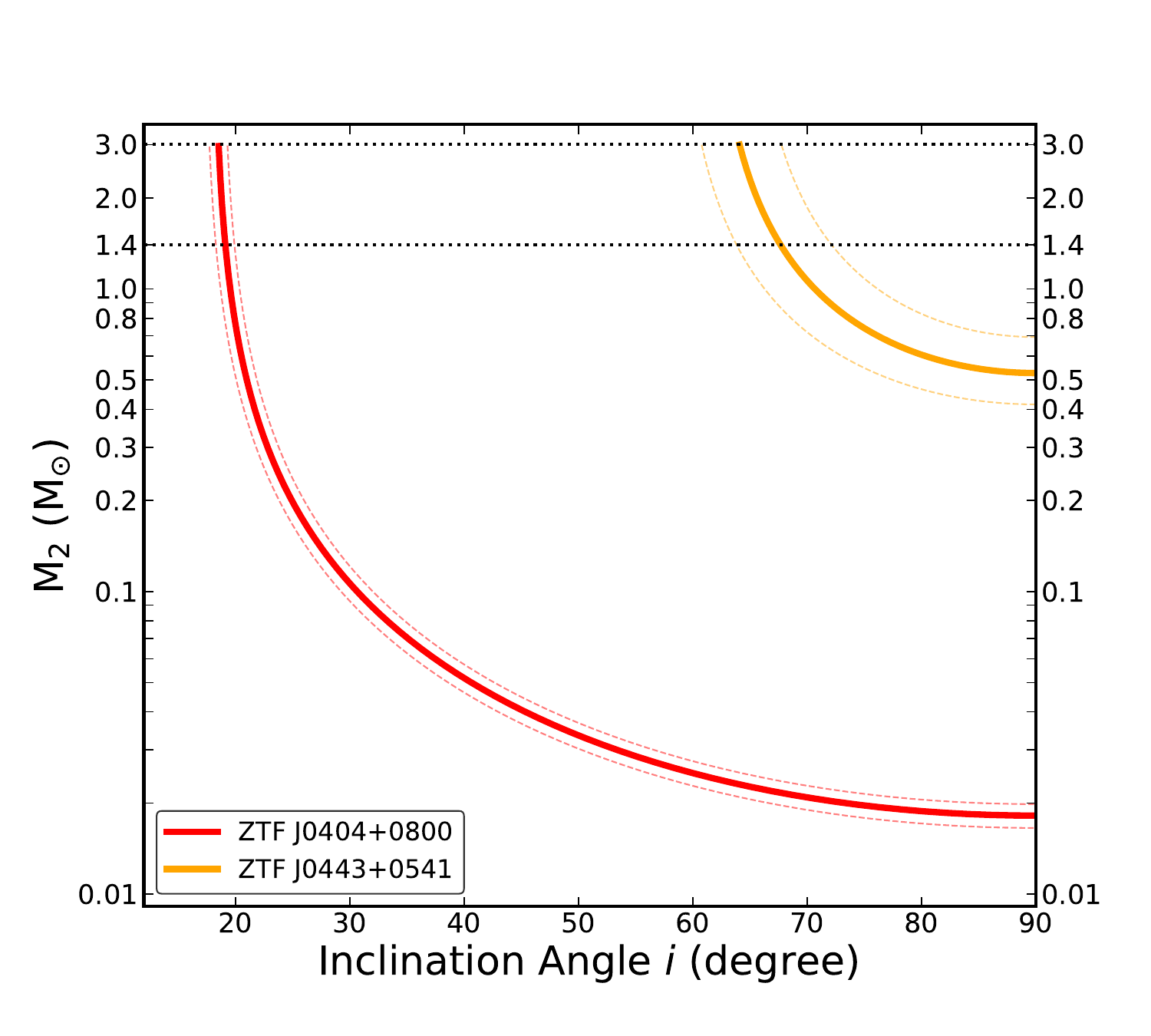}
            \caption{Companion mass $M_{2}$ versus orbital inclination $i$ for J0404+0800 and J0443+0541. The solid lines represent their relationships inferred from Equation (\ref{eq:ELL}), with other measured parameters fixed. The dashed lines depict the results when $A_{\rm EV}$ is increased or decreased by 1$\sigma$. Two black dotted lines indicate the isopleths for masses of 1.4 $M_{\odot}$ and 3 $M_{\odot}$.}
            \label{fig:M2VSinl}
        \end{figure}
        
        It is noteworthy that only about 0.45\% of simulations yield valid solutions for J0756+6704, which indicates the notable discrepancy between the measurements and the expected model. Nevertheless, we increase the number of MC simulations to 10$^7$ in order to obtain more available solutions to provide referable estimates.

        For the remaining two sources lacking mass functions, their relationships between companion mass and orbital inclination, as inferred from Equation (\ref{eq:ELL}), are displayed in Figure \ref{fig:M2VSinl}. It can be observed that J0404+0800 unlikely host a NS unless the inclination angle is less than $\sim$ 20$^{\circ}$. To give a referable value of $M_{2}$, we assume the orientation of the orbital angular momentum is arbitrary, equivalent to randomly sampling $\cos i$ between 0 and 1. Similarly, we employ 100,000 MC simulations to solve Equation (\ref{eq:ELL}) and accept the solutions that $M_{2} <3~ M_{\odot}$ and $M_{1} <$ 1.4 $M_{\odot}$. The final constrained parameters ($M_{1}$, $R_{1}$, $M_{2}$ and $i$) of nine selected targets can be seen in Table \ref{tab:estimateM2}. 

        \begin{table*}
        \tiny
        \centering
        \caption{Parameters Constrained from MC Simulations and Radio Observations for 9 Selected Targets}
        \begin{tabular}{lccccccccccc}
            \toprule
            Name & $M_1$ & $R_1$ & $M_2$ & $i$ & $A_{\rm EV}$ & $T_{\rm 0,sup\_conj}$ & $\mu$ & $\tau$ & $f_{1}(M_{2}$) &Radio & $L_{\rm 1400~MHz}$\\
            & ($M_{\odot}$) & $(R_{\odot}$) & ($M_{\odot}$) & ($^{\circ}$) & (\%) & (MJD) & & & ($M_{\odot}$) & & (mJy kpc$^{2}$) \\
            \hline
            ZTF J0101+0401 & 0.193$^{+0.011}_{-0.011}$ & 0.191$^{+0.013}_{-0.014}$ & 0.635$^{+0.987}_{-0.233}$ & 47.049$^{+23.971}_{-17.740}$ & 1.32(16) & 58281.49558 & 0.53 & 0.27 & 0.151(19) & FAST & < 0.0139 \\ 
            ZTF J0404+0800 & 0.172$^{+0.014}_{-0.014}$ & 0.222$^{+0.034}_{-0.030}$ & 0.027$^{+0.044}_{-0.013}$ & 61.829$^{+19.544}_{-24.274}$ & 1.11(09) & 58206.17723 &0.51 & 0.74 & - & FAST & < 0.0058 \\
            ZTF J0443+0541 & 0.160$^{+0.013}_{-0.013}$ & 0.224$^{+0.023}_{-0.018}$ & 0.421$^{+0.697}_{-0.227}$ & 74.819$^{+10.491}_{-13.938}$ & 2.69(16) & 58334.53492 &0.54 & 0.27 & - & FAST & < 0.0156 \\
            ZTF J0745+1949 & 0.160$^{+0.010}_{-0.010}$ & 0.056$^{+0.005}_{-0.005}$ & 0.280$^{+0.282}_{-0.116}$ & 30.378$^{+15.096}_{-8.428}$ & 1.61(09) & 58205.18216 & 0.55 & 0.30 & 0.017(3) & FAST & < 0.0003 \\
            ZTF J0756+6704 & 0.181$^{+0.011}_{-0.008}$ & 0.321$^{+0.036}_{-0.096}$ & 1.326$^{+1.190}_{-0.437}$ & 54.169$^{+27.036}_{-15.451}$ & 0.90(08) & 58207.49116 & 0.48 & 0.70 & 0.546(33) & GBT & < 0.2550 \\
            ZTF J1048-0000 & 0.191$^{+0.010}_{-0.020}$ & 0.087$^{+0.007}_{-0.007}$ & 0.679$^{+0.179}_{-0.032}$ & 76.152$^{+9.840}_{-16.545}$ & 8.42(78) & 58199.31062 & 0.55 & 0.29 & 0.383(31) & FAST & < 0.0039 \\
            ZTF J1238+1946 & 0.195$^{+0.012}_{-0.012}$ & 0.260$^{+0.020}_{-0.014}$ & 1.327$^{+0.695}_{-0.330}$ & 50.050$^{+9.887}_{-9.870}$ & 2.44(14) & 58202.43981 & 0.42 & 0.58 & 0.453(14) & GBT & < 0.0552 \\
            ZTF J1401-0817 & 0.269$^{+0.009}_{-0.005}$ & 0.117$^{+0.008}_{-0.010}$ & 1.167$^{+0.902}_{-0.297}$ & 58.778$^{+20.141}_{-17.158}$ & 10.37(16) & 58203.46077 & 0.53 & 0.28 & 0.487(12) & FAST & < 0.0024 \\ 
            ZTF J1741+6526 & 0.174$^{+0.003}_{-0.002}$ & 0.064$^{+0.004}_{-0.003}$ & 1.150$^{+0.190}_{-0.049}$ & 78.113$^{+9.248}_{-12.204}$ & 1.42(11) & 58197.39962 &0.50 & 0.75 & 0.801(39) & GBT & < 0.0908 \\ 
            \bottomrule
        \end{tabular}
        \raggedright\footnotesize{\textbf{Note.} Four main parameters ($M_{1}$, $R_{1}$, $M_{2}$ and $i$) are constrained from MC simulations, as described in Section \ref{subsec: M2}. $A_{\rm EV}$  is the amplitude of ellipsoidal variation separated from ZTF g-band lightcurves. $T_{\rm 0,sup\_conj}$ is the referenced epoch corresponding to superior conjunction of the ELM WD. $\mu_{1}$ and $\tau_{1}$ are linear limb-darkening coefficient and gravity-darkening coefficient used in MC simulations. Mass function $f_{1}(M_{2}$) is calculated from Equation (\ref{eq:mass_func}), and the related parameters are provided from the Ref. listed in Table \ref{tab:ELLtarget}. The ``Radio'' column indicates the telescopes used for radio searches. Specifically, six sources were observed with FAST in this work, while three sources were observed using the GBT with no radio pulsations found \citep{2020ApJ...889...49B,2021MNRAS.505.4981A}. $L_{\rm 1400~MHz}$ represents the upper limit of radio luminosity at 1,400 MHz.}
        \label{tab:estimateM2}
        \end{table*}        

        The estimated parameters of J1741+6526 and J0745+1949 can be compared with the values reported in \cite{2014ApJ...792...39H}. We found the results of J1741+6526 agree well with the reported values. However, considerable disagreements emerge in the outcomes for J0745+1949. Specifically, our estimations for $R_{1}$ and $i$ are notably smaller than the reported values ($R_{1}\sim0.176~R_{\odot}$, $i\sim63.2^{\circ}$). In fact, our results are much closer to the values ($R_{1}\sim0.057~R_{\odot}$, $i\sim10.5^{\circ}$) inferred from \cite{2018arXiv180905623B}.

        In our estimations, the companion masses of 4 targets exceed 1 $M_{\odot}$, making them more likely candidates for harboring MSPs. However, for the other sources, the possibility of hosting MSPs still exists as long as the orbital inclination is smaller than expected. Follow-up radio or X-ray observations around ELM WDs can help confirm or eliminate the existence of MSPs (e.g. \citealt{2009ApJ...697..283A,2020ApJ...889...49B}).
    
    \subsection{Radio Pulse Search} \label{sec:mwc}
        Compared to normal pulsars, MSPs generally have larger radio beaming fractions, e.g., from 50\% to 90\% \citep{1998ApJ...501..270K}, resulting in the high chance of detecting radio pulsations. In this study, we select 6 targets from Table \ref{tab:estimateM2} that have not yet been observed in radio, and perform radio observations for them by using Five-hundred-meter Aperture Spherical radio Telescope (FAST) in 2023 (PID: PT2023\_0058), to search for potential pulse signals. Each target was carried out a 15-min L-band tracking observation, with the first minute dedicated to injecting modulated noise to calibrate the polarized pulse signal once the radio pulsations are discovered. The sampling time was chosen to be $\sim$50 $\mu$s, and the number of frequency channels was set to 4096. 
    
        The observational data were processed by {\ttfamily PRESTO}\footnote{\url{https://github.com/scottransom/presto}} \citep{2002AJ....124.1788R}, one of classic pulsar search softwares. Firstly, the radio frequency interference (RFI) were automatically eliminated, and then the data were de-dispered with the trial dispersion measures (DM) ranging from 0 to 1000 pc cm$^{-3}$. Next the time-domain signals were transformed into frequency-domain signals via fast Fourier transform (FFT). Afterwards, the acceleration searches were executed with {\ttfamily $z_{\rm max}$} (the maximum Fourier frequency derivative) of 200 to enhance sensitivity towards short-orbit systems.
        
        However, no convincing radio pulse signal was found among these targets. The sensitivity can be estimated by radiometer equation:
        \begin{equation}
            S_{\rm min} = \frac{{\rm (S/N)_{min}} (T_{\rm sys}+T_{\rm sky})}{G\sqrt{n_{\rm p}t_{\rm int}\Delta f}} \times \sqrt{\frac{\delta}{1-\delta}},
        \end{equation}
        where ${\rm (S/N)_{min}}$ is threshold of signal-to-noise ratio, $T_{\rm sys}$ is the system temperature, $T_{\rm sky}$ is the sky background temperature, $G$ is the antenna gain, $T_{\rm int}$ is the integration time, $n_{p}$ is the number of polarizations, $\Delta f$ is the observed bandwidth and $\delta$ is the duty cycle of pulse. 
    
        For FAST L-band observations with central frequency of 1250 MHz, $T_{\rm sys}$=24 K, $G$=16 K Jy$^{-1}$ and $n_{p}$=2. The actual integration time $T_{\rm int}$ of each target is 840 s. The effective bandwidth $\Delta f$ is 400 MHz. The sky background temperature $T_{\rm sky}$ at 1,400 MHz is inferred from the 408-MHZ All-Sky Continuum Survey \citep{1982A&AS...47....1H}.
        To evaluate $\delta$, we utilize the mean value (0.29) of $W_{10}/P_{\rm spin}$ for MSPs orbited by He WDs, collected from ATNF V1.70 catalogue\footnote{\url{https://www.atnf.csiro.au/research/pulsar/psrcat/}}\citep{2005AJ....129.1993M}, where $W_{10}$ represents the pulse width at 10\% level of the peak and $P_{\rm spin}$ is the spin period. Applying the parameters mentioned above, the upper limit of radio flux $S_{\rm min}$ at 1400 MHz can be estimated to be $\sim$ 8 $\mu$Jy when the expected ${\rm (S/N)_{min}}$ is 7 and the spectral index is assumed to be $-$1.4 \citep{2013MNRAS.431.1352B}. Then, the upper limit of pseudo luminosity, $L_{\rm 1400~MHz}$, can be calculated by $S_{\rm min}d^{2}$, where $d$ is the spectrophotometric distance. It is noteworthy that the non-detection of radio pulsations could also be attributed to some factors, such as the emission beam failing to point towards the Earth, or the radio signals being obscured by the companion star. 
        
        On the other hand, MSPs also exhibit abundant thermal or non-thermal X-ray emissions, originating from the heated polar cap, magnetosphere and intrabinary shock, and their X-ray luminosity has been found to be positively correlated with spin-down luminosity \citep{2018ApJ...864...23L,2023ApJ...944..225L}.
        We have also cross-matched our targets with X-ray catalogues, such as Swift 2SXPS\footnote{\url{https://www.swift.ac.uk/2SXPS/}}, 4XMM DR13\footnote{\url{http://xmmssc.irap.omp.eu/Catalogue/4XMM-DR13/4XMM\_DR13.html}} and Chandard CSC 2.1\footnote{\url{https://cxc.cfa.harvard.edu/csc/}}, within a search radius of 1 arcminutes. Still, no X-ray counterpart was found. Combining the results of radio and X-ray searches, these six targets seem to resemble double WDs rather than MSP/WD systems.

\section{Constraining the Fraction of Ellipsoidal ELM-WD/MSP Systems} \label{sec:discussion}
    Many attempts to search for MSPs around low-mass WDs have been unsuccessful, supporting the view that the fraction of MSPs hosting in low-mass WD binary systems should be intrinsically low. \cite{2001MNRAS.324..797S} predict the birth rate of double WDs is two orders of magnitude higher than that of NS/WD systems. In fact, a considerable fraction of pulsar binaries may be disrupted during supernova explosions due to excessive mass loss or high kick velocities. \cite{2014ApJ...797L..32A} suggested that most low-mass WDs are accompanied by CO-core WD companions by using a mixture Gaussians model, and the NS companion fraction $f_{\rm NS}$ is estimated to be $<$ 16\% within 1$\sigma$ uncertainty. \cite{2021MNRAS.505.4981A} inferred that the $f_{\rm NS}$ is even less than 10\%, based on the non-detection of radio signals for a group of low-mass WD binaries. 

    In this study, our primary focus is on examining the fraction ($f_{\rm MSP,ELL}$) of ELM WDs with observable ellipsoidal variations orbited by MSP companions, based on the fact that no radio pulsations were detected. Similar to the method described in \cite{2007MNRAS.374.1437V} and \cite{2009ApJ...697..283A}, the expected $f_{\rm MSP,ELL}$ can be simply estimated by using
        \begin{equation}
            \prod_{j = 1}^{N} (1 - P_{\rm beam}\times P_{L,j} \times P_{\rm eff} \times f_{\rm MSP, ELL}) > \frac{1}{2},
        \end{equation}
    where $P_{\rm beam}$ is the radio beaming fraction, $P_{L,j}$ represents the radio luminosity completeness of $j^{\rm th}$ source, and $P_{\rm eff}$ denotes the success rate of pulsar search algorithm.

    We assume $P_{\rm beam}$ of MSPs to be 0.7$\pm$0.2 \citep{1998ApJ...501..270K}, and $P_{\rm eff}$ is set to a modest value of 0.8. To determine the $P_{L,j}$, the underlying luminosity distribution of MSPs should be known at first, which is still unclear at present. Here we employ the log-normal luminosity distribution with $\mu$=$-$1.1 and $\sigma$=0.9 \citep{2006ApJ...643..332F}, which are widely applied in population synthesis study of pulsars. Then each $P_{L,j}$ can be calculated as the proportion of luminosity exceeding $L_{\rm 1400~MHz}$ within the underlying luminosity distribution.

    There are 9 sources listed in Table \ref{tab:estimateM2}, which have been observed by FAST or Green Bank Telescope (GBT), enabling the calculation of $f_{\rm MSP,ELL}$. In addition, we incorporate two sources, J0056-0611 and J0112+1835, which exhibit ellipsoidal variation\footnote{Note that their ellipsoidal visibility was not found in the ZTF data.} and were observed by GBT \citep{2018arXiv180905623B,2020ApJ...889...49B}. Their $L_{\rm 1400~MHz}$ are inferred to be $<$ 0.0552 mJy kpc$^{2}$ by using the assumed spectral index of $-1.4$. Combining the results of 11 ellipsoidal ELM-WDs, the estimated $f_{\rm MSP,ELL}$ is below 15$^{+6}_{-3}$\%. It is worth noting that we did not consider the effects of assumed spin periods and DM on sensitivity\citep{2009ApJ...697..283A,2021MNRAS.505.4981A}, which could lead to an underestimation of the $f_{\rm MSP,ELL}$. On the other hand, if the luminosity of the MSP population is dimmer than the assumed distribution (e.g. \citealt{2020ApJ...905..144H}), or if the adopted spectrophotometric distances for some targets (e.g., J1048-0000, 1401-0817, and J0745+1949) are underestimated, leading to an overestimation of radio luminosity completeness, the $f_{\rm MSP,ELL}$ will also be underestimated. It is expected that the $f_{\rm MSP,ELL}$ is slightly larger than $f_{\rm NS}$, as ELM WDs are more easily tidally distorted by heavier MSPs than CO-core WDs in similar orbits, thereby exhibiting ellipsoidal variations. However, if the intrinsic $f_{\rm NS}$ is sufficiently small, the difference may be negligible. Incorporating more ellipsoidal ELM WDs would provide better constraints on $f_{\rm MSP,ELL}$.

\section{Conclusion} \label{sec:conclusion}
    ELM WDs may be tidally distorted by their heavier companions (e.g., MSPs or CO-core WD), leading to discernible ellipsoidal variations in time-series data. Identifying such systems can aid in the search for potential MSPs.
    In this study, we selected approximately 12,000 samples of ELM WDs and their candidates, and eventually identified 23 targets exhibiting ellipsoidal-like variations with orbital periods shorter than one day, using the public data from ZTF DR15. And 17 of them are newly discovered.
    
    We further selected nine high-priority targets that are likely consistent with the evolution of RL channel on the mass-period diagram as well as have well-measured atmospheric parameters. Their companion masses are estimated from the extracted ellipsoidal variation amplitude. According to our estimation, four of them have companion masses exceeding 1 $M_{\odot}$, making them more likely candidates for harboring MSPs. Although for other sources we cannot completely rule out the possibility that they host MSPs.
    
    The FAST observations were then performed on six targets of them to search for their potential radio pulsations, but no convincing pulse signals were detected. Their upper limit of radio flux at 1,400 MHz is about 8 $\mu$Jy, estimated using the radiometer equation. Combining the fact that their counterparts were not found in some X-ray catalogues, these systems are more likely to be double WDs. 
    Based on the non-detection of radio pulsations by GBT or FAST from 11 similar systems, the fraction of ellipsoidal ELM WDs around MSPs ($f_{\rm MSP,ELL}$) is estimated to be $<$ 15$^{+6}_{-3}$\%. 
    
    We anticipate that further multi-wavelength observations and studies of remaining ELL-type targets in this work, as well as new discovered ellipsoidal ELM WDs from other researches (e.g. \citealt{2023ApJS..264...39R}), can increase the chances of finding MSPs or better constrain the fraction of ellipsoidal ELM WDs around MSPs. Besides, about 4,300 candidates from our initial sample have a declination $<$ $-30^{\circ}$ and are not covered by the ZTF's footprint. These sources are expected to be searched for periodic signals using the the southern telescopes, such as the Large Synoptic Survey Telescope (LSST), in the future.  

\section*{Acknowledgements}
    We acknowledge the use of public lightcurve data from ZTF obtained through IRSA ZTF-LC-API queries (\url{https://irsa.ipac.caltech.edu/docs/program\_interface/ztf\_lightcurve\_api.html}) and public astrometric parameters from Gaia archive (\url{https://gea.esac.esa.int/archive/}). The VizieR catalogues and the ATNF pulsar catalogue were also utilized for this work. 
    This work made use of the data from FAST (Five-hundred-meter Aperture Spherical radio Telescope) (\url{https://cstr.cn/31116.02.FAST}). FAST is a Chinese national mega-science facility, operated by National Astronomical Observatories, Chinese Academy of Sciences. WJH and PHT thank the support from the National Natural Science Foundation of China (NSFC) under grant 12273122. LLR and JML gratefully acknowledge support from the NSFC through grant 12233013.
    
\vspace{5mm}
\software{{\ttfamily Astropy} \citep{2013A&A...558A..33A,2018AJ....156..123A}, {\ttfamily PRESTO} \citep{2002AJ....124.1788R}}

\bibliography{references}{}
\bibliographystyle{aasjournal}

\end{document}